\documentclass[plain]{aiaa-pretty}
\usepackage{amsthm}
\usepackage{amsmath}
\usepackage{amsfonts}
\usepackage{amssymb}
\usepackage{dsfont}
\usepackage{subfigure}
\usepackage{psfrag}
\usepackage{balance}
\usepackage{float}
\usepackage[titletoc,title]{appendix}

\let\doi\relax
\usepackage{doi}
\author{Mirko Leomanni\thanks{Research Associate, Department of Information Engineering and Mathematics, \texttt{$\!$leomanni@diism.unisi.it}},
Andrea Garulli \thanks{Professor, Department of Information Engineering and Mathematics, \texttt{garulli@diism.unisi.it}},
Antonio Giannitrapani \thanks{Professor, Department of Information Engineering and Mathematics, \texttt{giannitrapani@diism.unisi.it}},
Renato Quartullo \thanks{PhD Student, Department of Information Engineering and Mathematics, \texttt{quartullo@diism.unisi.it}}\\[1mm]
\textit{Universit\`a di Siena, Siena, 53100, Italy}}

\title{Satellite Relative Motion Modeling and Estimation\\ via Nodal Elements}

\abstract{In this paper, a new parametrization of the relative motion between two satellites orbiting a central body is presented. The parametrization is based on the nodal elements: a set of angles describing the orbit geometry with respect to the relative line of nodes. These are combined with classical orbital elements to yield a nonsingular relative motion description. The exact nonlinear, perturbed dynamic model resulting from the new parametrization is established. The proposed parameter set captures the fundamental Keplerian invariants, while retaining a simple relationship with local orbital coordinates. An angles-only relative navigation filter and a collision avoidance scheme are devised by exploiting these features. The navigation solution is validated on a case study of an asteroid flyby mission. It is shown that a collision can be detected early on in the estimation process, which allows one to issue a timely evasive maneuver.}

\begin{document}
\maketitle
\section{Introduction}\label{Intro}
Relative motion modeling has been widely investigated since the early space era \cite{clohessy1960terminal,lawden1963optimal,hempel1965rendezvous}. Nowadays it is receiving renewed attention, due to its pivotal role in a number of emerging multi-satellite applications. Examples include the Mars Sample Return \cite{mattingly2011} and the Proba-3 \cite{landgraf2013} missions, as well as comet/asteroid rendezvous missions like Rosetta \cite{Glassmeier2007}, Hayabusa 2 \cite{TSUDA2013356} and OSIRIS-REx \cite{Lauretta2017}. Broadly speaking, existing modeling approaches can be classified according to the state representation they use: relative inertial position and velocity; local relative position and velocity; relative orbital elements (the reader is referred to \cite{alfriend05,sullivan2017comprehensive} for a comprehensive overview). These parameterizations lead to different dynamic models. Each has its advantages and limitations, that must be carefully evaluated against the specific application requirements. For instance, in relative navigation it is desirable to suitably trade-off the complexity of the dynamic model and that of the output mapping, while for control purposes simple dynamic models are generally more convenient, see, e.g., \cite{Weiss12,quadrelli15,dicairano18,ManciniCSL20}.

Orbital element errors (OEE), either orbital element differences (OED \cite{schaubrelative,gim2005satellite}) or relative orbital elements (ROE \cite{montenbruck2006,d2006proximity,d2010autonomous}), offer a clear description of the orbit geometry through the fundamental Keplerian invariants. Taking advantage of this feature, analytical solutions for the relative motion problem have been developed in \cite{schaubrelative,gim2005satellite,vadali2002analytical}. References \cite{montenbruck2006,d2006proximity} present impulsive maneuvering strategies able to minimize collision risk for spacecraft in close formation.
Feedback solutions to the continuous-thrust orbital transfer and rendezvous control problems have been proposed in \cite{gurfil03,hartley2012695,leomanni2017class,leomanni2018state}. State estimation with ROE has been investigated in \cite{Gaias14,sullivan17}. In most of these works, the relative motion is linearized about a circular or an elliptical reference orbit.

In spite of their advantages, it has been noticed that OEE parametrizations provide an indirect representation of the relative motion problem \cite{kasdin2005canonical}. Specifically, their definition relies on the basic understanding that both the reference and the actual trajectories are expressed in an inertial coordinate system. This leads to a non-minimal description of the relative motion geometry. Indeed, one can show that a minimal parametrization of the relative orientation of two orbits requires to adopt a satellite-based reference system which is, in general, time-varying \cite[Sec.\ 1.4]{vallado2001fundamentals}. Very few methods have been proposed in the literature to address this issue, while retaining the key geometrical properties of OEE. In \cite{kasdin2005canonical}, a set of epicyclic elements is derived by exploiting local orbital coordinates. The approach is valid for small deviations about a circular orbit. In \cite{gurfilmanifold} and \cite[Sec.\ 7.2]{alfriend2009spacecraft}, a combination of Euler parameters and local translational states is used to tackle the full nonlinear problem. In the resulting parametrization, information on some of the Keplerian invariants is lost.

In this paper, the relative motion between two satellites in arbitrary elliptical orbits is described by introducing a new set of relative states. These are obtained from a suitable combination of the orbit shape parameters (eccentricity, semi-axis), of the inclination angle between the orbital planes, and of four angles describing the satellite instantaneous position and periapsis location with respect to the relative line of nodes, which we call \emph{nodal elements}. Such an element set allows one to remove redundant orbital parameters from the modeling problem and provides a straightforward characterization of intersecting orbits. The resulting parametrization overcomes the singularity of the relative elements described in \cite{gurfilmanifold}, for circular and coplanar orbits. With respect to the quaternion-based method in \cite{alfriend2009spacecraft}, the proposed approach leads to a lower-dimensional dynamic model. Moreover, it retains a direct relationship with both the Keplerian invariants and local translational states.

A collision detection technique and an angles-only relative navigation scheme are devised by exploiting these features. The collision detection problem is tackled by suitably extending the concept of passive safety introduced in \cite{d2006proximity} to the case of elliptical orbits. An extended Kalman filter (EKF) processing azimuth, elevation, and angular size measurements is employed for state estimation (see, e.g., \cite{Kim07,woffinden07}). The filter design is streamlined by the availability of a compact mapping between the new parameter set and the measured outputs.

The proposed techniques are validated on a case study inspired by the Rosetta flyby of asteroid Lutetia. The results of numerical simulations show that the EKF is able to estimate the relative motion with good accuracy. Moreover, it is shown that the information provided by the filter allows one to detect a potential collision well in advance. An optimal single-impulse collision avoidance maneuver based on this method is described.

The rest of the paper is organized as follows. Section \ref{RMD} introduces the proposed parametrization and illustrates its main geometrical features. In Section \ref{rmotmod}, the parameter dynamics are derived. The collision detection and EKF schemes are presented in Sections \ref{codect} and \ref{aanavigation}, respectively. They are validated on the considered flyby scenario in Section \ref{simb}. The main findings of this work are summarized in Section~\ref{conclu}.

\section{Relative Motion Description}\label{RMD}
In this section, some preliminary definitions are given and the new parametrization of the satellite relative motion is presented.

\subsection{Problem Setting}\label{prework}
Let us denote the central (attracting) body as the \emph{primary}, and refer to two satellites orbiting the primary as satellite $j$, $j=1,2$. We consider a primary-centered inertial (PCI), a satellite-based perifocal (PQW), and a radial-transverse-normal (RTN) frame. By convention, the fundamental plane of the PCI frame is taken coincident with the equatorial plane of the primary. The PCI frame axes are denoted by $X_{I}$, $Y_{I}$ and $Z_{I}$. The PQW frame of satellite $j$ (denoted by PQW$_j$) is a coordinate system centered at the primary, whose $X_{j}$ and $Z_{j}$ axes are aligned with the orbit periapsis and the orbit normal of satellite $j$, respectively, while the $Y_j$ axis completes a right handed triad. The PQW frame is, in general, time-varying, because the orbital plane varies in response to perturbations. The RTN frame of satellite $j$ (denoted by RTN$_j$) is obtained by rotating the PQW$_j$ frame by an angle equal to the true anomaly of satellite $j$ about the $Z_{j}$ axis and translating the origin to the satellite's center of mass.

\begin{figure}[!t]
\centering
\psfrag{x}{$X_{I}$}\psfrag{q}{$Z_{I}$}\psfrag{r}{$X_{2}$}\psfrag{t}{$X_{1}$}\psfrag{a}{$\alpha_2$}\psfrag{b}{$\alpha_1$}\psfrag{g}{$\gamma$}\psfrag{i}{$i_1$}\psfrag{k}{$i_2$}\psfrag{w}{$\Omega_1$}\psfrag{z}{$\Omega_2$}\psfrag{o}{$\omega_1$}\psfrag{p}{$\omega_2$}
\includegraphics[width=0.4\textwidth]{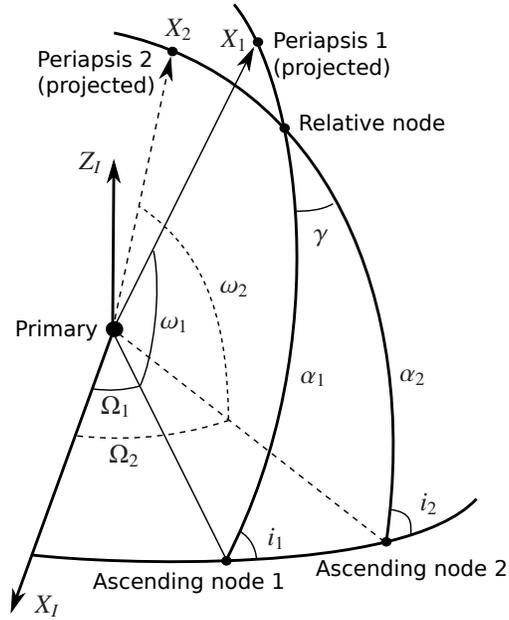}   %submit
\caption{Orbit geometry on the unit sphere, with respect to the PCI frame.}\label{relel}
\end{figure}

A standard representation of the orbit of the two satellites in the PCI frame is given by the classical orbital elements $\{a_j,e_j,i_j,\Omega_j,\omega_j,v_j\}$, $j=1,2$, namely the semimajor axis, eccentricity, inclination, right ascension of the ascending node, argument of perigee, and true anomaly. The elements $a_j$, $e_j$ and $v_j$ describe the motion of satellite $j$ in the $X_{j} Y_{j}$-plane. The Euler angles $i_j,\Omega_j,\omega_j$ describe the orientation of the PQW$_j$ frame relative to the PCI frame.  The direction cosine matrix which transforms the PCI frame to the PQW$_j$ frame is given by \cite{vallado2001fundamentals}
\begin{equation}\label{pqwmap}
\mathbf{T}_{\text{PCI}}^{\text{PQW}_j}=\mathbf{T}_Z(\omega_j)\mathbf{T}_X(i_j)\mathbf{T}_Z(\Omega_j)
\end{equation}
where $\mathbf{T}_X(\theta)$ and $\mathbf{T}_Z(\theta)$ are elementary rotations of the coordinate basis vectors by an angle $\theta$ about the $X$ and $Z$ axes, respectively. The geometry of the rotation sequence is illustrated in Fig.~\ref{relel}, which depicts the intersections of the fundamental plane of the inertial frame, and of the two orbital planes of satellites 1 and 2, with the unit sphere.

The relative orientation between the two satellite orbits is usually specified by the direction cosine matrix $\mathbf{T}_{\text{PQW}_2}^{\text{PQW}_1}$, which brings the $X_{2}$, $Y_{2}$ and $Z_{2}$ axes to the $X_{1}$, $Y_{1}$ and $Z_{1}$ axes. According to \eqref{pqwmap}, it can be expressed as
\begin{equation}\label{rpqwmap}
\begin{array}{lll}
\mathbf{T}_{\text{PQW}_2}^{\text{PQW}_1}&=&\mathbf{T}_{\text{PCI}}^{\text{PQW}_1}\left(\mathbf{T}_{\text{PCI}}^{\text{PQW}_2}\right)^T\\
&=&\mathbf{T}_Z(\omega_1)\mathbf{T}_X(i_1)\mathbf{T}_Z(\Omega_1-\Omega_2)\mathbf{T}_X(-i_2)\mathbf{T}_Z(-\omega_2),
\end{array}
\end{equation}
where the superscript $T$ denotes the transpose operator. Notice that it is always possible to parameterize $\mathbf{T}_{\text{PQW}_2}^{\text{PQW}_1}$ via a minimal sequence of three elementary rotations. However, the use of the PCI frame as an intermediate frame in the right hand side of \eqref{rpqwmap} results in a larger number of rotations, due to the lack of commutativity of the rotation sequence. For this reason, modeling approaches based on the PCI frame, such as OEE, cannot provide a minimal parametrization of the relative motion.

\begin{figure}[t]
\centering
\psfrag{a}{$\theta_2$}\psfrag{b}{$\theta_1$}\psfrag{g}{$\gamma$}\psfrag{q}{$\lambda_2$}\psfrag{o}{$\psi$}\psfrag{p}{$\lambda_1$}
\includegraphics[width=0.4\textwidth]{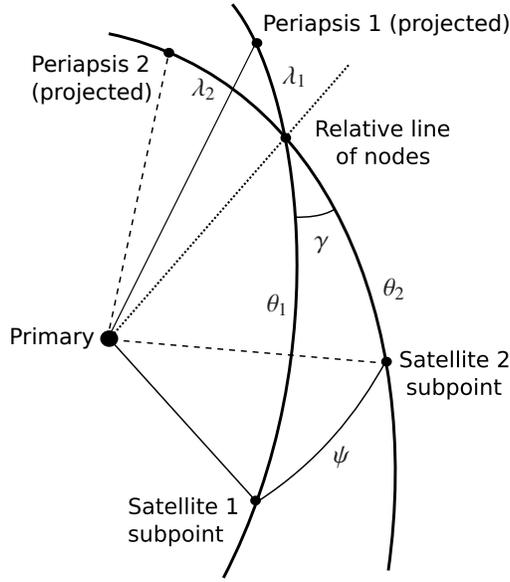}   %submit
\caption{Orbit geometry on the unit sphere, with respect to the relative node.}\label{relel2}
\end{figure}
As outlined in \cite{gurfilmanifold}, the above shortcoming can be amended by introducing a set of relative orientation parameters describing the transformation $\mathbf{T}_{\text{PQW}_2}^{\text{PQW}_1}$. In particular, let us define the \emph{relative line of nodes} as the intersection of the orbital planes of satellites 1 and 2 (see Fig.~\ref{relel2}). We refer to the \emph{relative node} as the projection on the unit sphere of the point at which spacecraft 2 crosses the orbital plane of spacecraft 1, in the ascending direction. Let $\gamma$ be the angle between the two orbital planes (relative inclination), and $\lambda_j$, $j=1,2$, be the angles made by the periapses of the two satellites with respect to the relative line of nodes. Then, it is not difficult to see that
\begin{equation}\label{rpqwmap1}
\mathbf{T}_{\text{PQW}_2}^{\text{PQW}_1}=\mathbf{T}_Z(\lambda_1)\mathbf{T}_X(-\gamma)\mathbf{T}_Z(-\lambda_2),
\end{equation}
and, in particular, that the rotation sequence parameterized by the Euler angles $\lambda_j$ and $\gamma$ is of minimal length.
Despite achieving minimality, this representation raises two fundamental concerns. First, $\lambda_j$ is undefined when orbit $j$ is circular. As a consequence, the transformation \eqref{rpqwmap1} is singular if $e_j=0$ for some $j$. Moreover, the relative line of nodes is undefined for coplanar orbits ($\gamma=0,\,\pi$). The singularity at $\gamma=0$ is particularly annoying, since this configuration plays a key role in several applications (e.g., coplanar rendezvous).

In the following, a new modeling approach is presented to address the aforementioned minimality and singularity issues. The main idea behind our approach is to extend \eqref{rpqwmap1} so as to express the relative orientation between the $\text{RTN}_2$ and $\text{RTN}_1$ frames, according to the transformation
\begin{equation}\label{rpqwmap2}
\mathbf{T}_{\text{RTN}_2}^{\text{RTN}_1}=\mathbf{T}_Z(v_1+\lambda_1)\mathbf{T}_X(-\gamma)\mathbf{T}_Z(-\lambda_2-v_2),
\end{equation}
and to develop a nonsingular parametrization of the relative motion accounting for the rotations in the right hand side of~\eqref{rpqwmap2}.

\subsection{Proposed Parametrization}\label{propar}

In addition to the orbital parameters described above, let us consider the angles $\theta_j$, $j=1,2$, made by the two satellite position vectors with respect to the relative line of nodes (see Fig.~\ref{relel2}). Notice that these are fast periodic variables, whose period is equal to the orbital period. Contrary to the true anomalies $v_j$, they are nonsingular for circular orbits.  We call the parameters $\{\lambda_j,\theta_j\}$ \emph{nodal elements}, highlighting the fact that they all refer to the relative node. Similarly to the relative inclination angle $\gamma$, nodal elements are determined uniquely by the mutual configuration of the two orbits, and hence they do not require the definition of a PCI frame. This feature, together with their simple geometrical interpretation, makes them promising candidates to parameterize the relative motion. The relation between nodal and classical elements is given by
\begin{equation}\label{releo}
\begin{array}{lll}
\theta_j&=&v_j+\lambda_j  \\
\lambda_j&=&\omega_j-\alpha_j
\end{array}
\end{equation}
being $\alpha_j$ the angle between the relative node and the ascending node $j$. The first equation in \eqref{releo} allows one to rewrite \eqref{rpqwmap2} as
\begin{equation}\label{rpqwmap3}
\mathbf{T}_{\text{RTN}_2}^{\text{RTN}_1}=\mathbf{T}_Z(\theta_1)\mathbf{T}_X(-\gamma)\mathbf{T}_Z(-\theta_2).
\end{equation}
The angles $\alpha_j$  and $\gamma$ in \eqref{releo}-\eqref{rpqwmap3} can be expressed in terms of classical elements trough the identity (see Fig.~\ref{relel})
\begin{equation}\label{abmap}
\mathbf{T}_Z(-\alpha_1)\mathbf{T}_X(-\gamma)\mathbf{T}_Z(\alpha_2)=\mathbf{T}_X(i_1)\mathbf{T}_Z(\Omega_1-\Omega_2)\mathbf{T}_X(-i_2).
\end{equation}

By suitably combining the classical elements $\{a_j,e_j\}$, the relative inclination angle $\gamma$, and the nodal elements $\{\lambda_j,\theta_j\}$, the orbital motion of satellite 2 relative to a reference satellite 1 can be described by the following parameters
\begin{align}
\begin{array}{lll}
 \delta \theta &=&\theta_2-\theta_1\\
 \delta p  &=& (p_2 - p_1)/p_1\\
 \delta \xi_x &=& e_2\cos(\theta_1-\lambda_2)-e_1 \cos( \theta_1-\lambda_1)\\
 \delta \xi_y &=& e_2\sin(\theta_1-\lambda_2)-e_1\sin(\theta_1-\lambda_1)\\
 \delta h_x &=& \tan(\gamma/2)\cos\theta_1\\
 \delta h_y &=& \tan(\gamma/2)\sin\theta_1,
\label{mypar}
\end{array}
\end{align}
where $p_j=a_j(1-e_j^2)$ is the semiparameter of orbit $j$. The parametrization \eqref{mypar} provides a physically meaningful measure of the relative motion error. In fact, $\delta \theta$ encodes the phase error,  $\delta p$ describes the semiparameter error,  $\delta \xi_x$ and  $\delta \xi_y$ are both identically zero only if $e_2=e_1$ and $\lambda_2=\lambda_1$, while $\delta h_x$ and  $\delta h_y$ are both identically zero only if $\gamma=0$. Moreover, $\delta \theta$, $\delta h_x$ and $\delta h_y$ embed a minimal representation of $\mathbf{T}_{\text{RTN}_2}^{\text{RTN}_1}$, since the right hand side of \eqref{rpqwmap3} can be expressed in terms of these parameters. When all the relative states $\{\delta \theta,\delta p,\delta \xi_x ,\delta \xi_y,\delta h_x,\delta h_y\}$ are zero, the two satellites are guaranteed to follow exactly the same path. Together with the reference parameters $\{p_1,e_1\cos(v_1),e_1\sin(v_1)\}$ appearing on the right hand side of \eqref{mypar} (recall from \eqref{releo} that $\theta_1-\lambda_1=v_1$), these relative states allow one to fully characterize the orbit geometry. In particular, the complete element set $\{a_j,e_j,\gamma,\lambda_j,\theta_j\}$ can be recovered.

Concerning the domain of definition of the proposed parametrization, it is worth noticing that \eqref{mypar} is nonsingular for all closed orbit pairs, except for purely retrograde ones ($\gamma=\pi$). Specifically,  Fig.~\ref{relel2} reveals that as $\gamma$ tends to zero, $\delta\theta$ must converge to the angle $\psi$ between the position vectors of the two satellites. Formally, this can be shown by using the haversine formula
\begin{equation}\label{haver}
\text{hav}\, \psi=\text{hav}\, \delta\theta +\sin \theta_1\sin \theta_2\, \text{hav}\,\gamma,
\end{equation}
where $\text{hav}\, x=\sin^2(x/2)$, which confirms that $|\delta\theta|=|\psi|$ at $\gamma=0$. Moreover, one can verify that $\delta \xi_x$ and $\delta \xi_y$ are well-defined at $e_j=0$, since the trigonometric terms containing  $\theta_1-\lambda_2=v_2-\delta \theta$  and $\theta_1-\lambda_1=v_1$ in \eqref{mypar} are multiplied by $e_2$ and $e_1$, respectively. A singularity at $\gamma=\pi$ still remains due to the presence of the term  $\tan(\gamma/2)$ in the definition of $\delta h_x$ and  $\delta h_y$. However, such an orbital configuration rarely occurs in practice.

The proposed parametrization allows one to remove the dependence on the PCI frame, while avoiding singularities at $\gamma=0$ and $e_j=0$. The price to pay is that the relative states $\{\delta \xi_x,\delta \xi_y\}$ and $\{\delta h_x,\delta h_y\}$, encoding the eccentricity and inclination errors, are fast variables. Due to the particular structure of the parameterization, this represents only a minor drawback. In fact, the two vectors $[\delta \xi_x\;\delta \xi_y]^T$ and $[\delta h_x\;\delta h_y]^T$ retain a direct relationship with the so-called relative eccentricity vector $[\delta e_x\;\delta e_y]^T$ and relative inclination vector $[\delta i_x\;\delta i_y]^T$. Within the proposed framework, these can be modeled as
\begin{equation}
\left[
\begin{array}{c}
\delta e_x\\
\delta e_y
\end{array}
\right]
=
\left[
\begin{array}{c}
e_2\cos\lambda_2-e_1\cos\lambda_1\\
e_2\sin\lambda_2-e_1\sin \lambda_1
\end{array}
\right],
\quad\;\;
\left[
\begin{array}{c}
\delta i_x\\
\delta i_y
\end{array}
\right]
=
\left[
\begin{array}{c}
\tan (\gamma/2)\\
0
\end{array}
\right].
\label{eivecex}
\end{equation}
Both vectors are expressed in a planar coordinate system whose $X$-axis is aligned with the relative line of nodes. From \eqref{mypar} and \eqref{eivecex}, it follows that
\begin{equation}
\left[
\begin{array}{c}
\delta \xi_x\\
\delta \xi_y
\end{array}
\right]
=\mathbf{R}(\theta_1)
\left[
\begin{array}{r}
\delta e_x\\
-\delta e_y
\end{array}
\right],
\quad\;\;
\left[
\begin{array}{r}
\delta h_x\\
\delta h_y
\end{array}
\right]
=\mathbf{R}(\theta_1)
\left[
\begin{array}{r}
\delta i_x\\
\;\,\delta i_y
\end{array}
\right]\label{eivrel}
\end{equation}
being
\begin{equation}
\mathbf{R}(\theta_1)=\left[
\begin{array}{c c}
\cos(\theta_1)&-\sin(\theta_1)\\
\sin(\theta_1)&\;\,\cos(\theta_1)\\
\end{array}\right]
\label{rotmat}
\end{equation}
a standard rotation matrix in $\mathbb{R}^2$. Hence, one has the identities
\begin{equation}
\begin{split}
& \delta \xi=\sqrt{\delta \xi_x^2+\delta \xi_y^2}=\sqrt{\delta e_x^2+\delta e_y^2}\\
& \delta h=\sqrt{\delta h_x^2+\delta h_y^2}=\sqrt{\delta i_x^2+\delta i_y^2}.
\end{split}\label{eivec}
\end{equation}
where $\delta \xi$ and $\delta h$ denote the magnitude of the relative eccentricity and inclination vectors, respectively. The phase angle $\delta\phi$ between these vectors reads
\begin{equation}\label{eccphase}
\delta\phi=\text{atan}_2(\delta h_y,\, \delta h_x)-\text{atan}_2(\delta \xi_y,\,\delta \xi_x).
\end{equation}
In Section \ref{codect}, it will be shown that the scalar parameters $\{\delta \xi,\, \delta h ,\,\delta\phi\}$ given by  \eqref{eivec}-\eqref{eccphase} provide relevant information about the safety of the satellite relative motion.
\subsection{Mapping to Translational States}
The mapping between the considered parameter set and the relative position in local orbital coordinates is obtained by observing that the position vector $\delta \mathbf{r}$ of satellite 2 relative to satellite 1, seen from the RTN$_1$ frame, is given by
\begin{equation}\label{posmap}
\delta \mathbf{r}=\mathbf{T}_{\text{RTN}_2}^{\text{RTN}_1}
\left[
\begin{array}{c}
r_2\\0\\0
\end{array}
\right]
-
\left[
\begin{array}{c}
r_1\\0\\0
\end{array}
\right],
\end{equation}
where
\begin{equation}\label{radius1}
r_1=\frac{p_1}{1+e_1 \cos v_1},
\end{equation}
\begin{equation}\label{radius2}
r_2=\frac{p_1(1+\delta p)}{1\!+\!(\delta\xi_x\!+\!e_1\!\cos v_1)\cos\!\delta\theta-\!(\delta\xi_y\!+\!e_1\!\sin v_1)\sin\!\delta\theta}.
\end{equation}
By using \eqref{rpqwmap3} and \eqref{radius1}-\eqref{radius2}, the mapping \eqref{posmap} can be expressed as an explicit function of the relative states \eqref{mypar}, according to
\begin{equation}
\frac{\delta \mathbf{r}}{r_1}=q
\left[
\begin{array}{c}
b_1\\
b_2\\
b_3
\end{array}
\right]
-
\left[
\begin{array}{c}
1\\ 0\\ 0
\end{array}
\right],\label{orbmotion}
\end{equation}
where the vector $[b_1\;b_2\;b_3]^T$ has unitary norm and
\begin{equation}\label{posquotient}
\begin{array}{c l l}
q&=&\dfrac{r_2}{r_1}=\dfrac{(1+\delta p)(1+e_1 \cos v_1)}{1\!+\!(\delta\xi_x\!+\!e_1\!\cos v_1)\cos\!\delta\theta-\!(\delta\xi_y\!+\!e_1\!\sin v_1)\sin\!\delta\theta}\\[4mm]
b_1&=&\dfrac{(1+\delta h_x^2-\delta h_y^2)\cos \delta\theta-2 \delta h_x \delta h_y\sin \delta\theta}{1+\delta h_x^2+\delta h_y^2}\\[4mm]
b_2&=&\dfrac{(1-\delta h_x^2+\delta h_y^2)\sin \delta\theta-2 \delta h_x \delta h_y\cos \delta\theta}{1+\delta h_x^2+\delta h_y^2}\\[4mm]
b_3&=&\dfrac{2 \delta h_y\cos \delta\theta+2 \delta h_x\sin \delta\theta}{1+\delta h_x^2+\delta h_y^2}.
\end{array}
\end{equation}
%By using \eqref{mypar} and the fact that $\theta_1=v_1+\lambda_1$, the mapping \eqref{orbmotion}-\eqref{posquotient} can be expressed as a function of the variables $v_1\in(-\pi,\;\pi]$ and $\delta \theta\in(-\pi,\;\pi]$ %only, all the other parameters being constant.
When the reference orbit is circular, i.e., $e_1=0$, \eqref{posquotient} is determined uniquely by the six relative states \eqref{mypar}.
The satellite relative velocity can be obtained by differentiating \eqref{orbmotion}-\eqref{posquotient} with respect to time, as it will be shown in Section \ref{rmotmod}.

Compared to existing orbital-element-based descriptions, the parametrization \eqref{mypar} enjoys a simpler nonlinear mapping to local translational states. In particular, the mapping \eqref{orbmotion}-\eqref{posquotient} does not require the absolute (inertial) orientation of the reference orbit to be known. This feature is especially relevant for relative navigation applications, in which the satellite motion is typically measured in terms of RTN coordinates. Compared to translational states, the parameters \eqref{mypar} provide a deeper insight of the orbital geometry, thanks to their clear relationship with the relative motion invariants.

\section{Relative Motion Dynamics}\label{rmotmod}
In this section, the exact nonlinear motion equations governing \eqref{mypar} are derived and their solution is briefly discussed. First, we restrict our attention to the ideal Keplerian dynamics. The perturbed case is addressed in Section \ref{rmotmod}.\ref{pertdyn}.

\subsection{Unperturbed dynamics}
Let us collect the relative states \eqref{mypar} in the vector
\begin{equation}\label{errorstate}
\textbf{\oe}=[\delta \theta\;\; \delta p\;\; \delta \xi_x \;\; \delta \xi_y\;\; \delta h_y\;\;\delta h_y]^T
\end{equation}
and define the reference parameter vector
\begin{equation}\label{etavec}
\boldsymbol{\eta}=[p_1\;\;e_1\cos v_1\;\;e_1 \sin v_1]^T.
\end{equation}

By adopting the shorthand notation $\dot{x}=\text{d}x/\text{d}t$, the unperturbed dynamics of \eqref{errorstate} can be expressed as
\begin{equation}
\dot{\textbf{\oe}}=\mathbf{f}(\textbf{\oe};\boldsymbol{\eta})
\label{mypardyn}
\end{equation}
where
\begin{equation}
\mathbf{f}(\textbf{\oe};\boldsymbol{\eta})=
\dot{v}_1
\left[
\begin{array}{c}
\dfrac{\Big(1+(\delta\xi_x+e_1\cos v_1)\cos\delta\theta-(\delta\xi_y+e_1\sin v_1)\sin\delta\theta\,\Big)^2}{(1+e_1\cos v_1)^2 \sqrt{(1+\delta p)^3}}-1\\
\;\,\,0\\
-\,\delta \xi_y\\
\;\,\delta \xi_x\\
-\,\delta h_y\\
\;\,\,\delta h_x
\end{array}
\right]
\label{mypardyn2}
\end{equation}
\begin{equation}\label{dotv1}
\dot{v}_1=\sqrt{\frac{\mu}{p_1^3}}(1+e_1\cos v_1)^2
\end{equation}
and $\mu$ is the gravitational parameter.
The nonlinear system \eqref{mypardyn}-\eqref{dotv1} is time-varying for elliptical reference orbits ($e_1\neq 0$). It becomes autonomous for circular ones ($e_1=0$).

The first equation of system \eqref{mypardyn} lacks a closed-form solution. On the other hand, the state $\delta p$ is constant and the last four equations in \eqref{mypardyn} can be solved in terms of $v_1(t)$ to give
\begin{equation}
\left[
\begin{array}{c}
\delta \xi_x(t)
\\
\delta \xi_y(t)
\end{array}
\right]
=
\mathbf{R}(v_1(t)-v_1(t_0))
\left[
\begin{array}{c}
\delta \xi_x(t_0)
\\
\delta \xi_y(t_0)
\end{array}
\right]\label{sigmasol}
\end{equation}
\begin{equation}
\left[
\begin{array}{c}
\delta h_x(t)
\\
\delta h_y(t)
\end{array}
\right]
=
\mathbf{R}(v_1(t)-v_1(t_0))
\left[
\begin{array}{c}
\delta h_x(t_0)
\\
\delta h_y(t_0)
\end{array}
\right]\label{hsol}
\end{equation}
where $v_1(t_0)$ indicates the value of $v_1$ at the initial time $t_0$. Notice that $v_1(t)-v_1(t_0)=\theta_1(t)-\theta_1(t_0)$ for the unperturbed motion. Moreover, $v_1(t)-v_1(t_0)=(t-t_0)\sqrt{\mu/a_1^3}$ for circular reference orbits.

The unperturbed dynamics of the parameter vector $\boldsymbol{\eta}$ are obtained from \eqref{etavec} and \eqref{dotv1} as
\begin{equation}\label{uvdyn}
\dot{\boldsymbol{\eta}}=\mathbf{f}_\eta(\boldsymbol{\eta})
\end{equation}
where
\begin{equation}
\mathbf{f}_\eta(\boldsymbol{\eta})=
\sqrt{\frac{\mu}{p_1^3}}(1+e_1\cos v_1)^2
\left[
\begin{array}{c}
0\\
-\,e_1\sin(v_1)\\
\;\,\,e_1\cos(v_1)
\end{array}
\right]
\label{uvdyn2}
\end{equation}
The satellite relative velocity, expressed in the RTN$_1$ frame, can be found from  \eqref{orbmotion}-\eqref{dotv1} and \eqref{uvdyn}-\eqref{uvdyn2}, according to
\begin{equation}\label{vellmap}
\delta \dot{\mathbf{r}}=\frac{\partial (\delta \mathbf{r})}{\partial\,\textbf{\oe}}\, \mathbf{f}(\textbf{\oe};\boldsymbol{\eta})+\frac{\partial (\delta \mathbf{r})}{\partial\, \boldsymbol{\eta}}\,\mathbf{f}_\eta(\boldsymbol{\eta}).
\end{equation}

\subsection{Perturbed dynamics}\label{pertdyn}
Let ${\mathbf{u}}_1$ and ${\mathbf{u}}_2$ be the perturbing accelerations acting on satellite 1 and 2, respectively, expressed in their own RTN frame.
In order to ease the derivation of the perturbed dynamics, define
\begin{equation}\label{tildeu}
\begin{array}{l l l l l}
\tilde{\mathbf{u}}_1&=&[\tilde{u}_{R1}\;\tilde{u}_{T1}\;\tilde{u}_{N1}]^T&=&\dfrac{r_1}{\sqrt{\mu p_1}}\,{\mathbf{u}}_1\\[2mm]
\tilde{\mathbf{u}}_2&=&[\tilde{u}_{R2}\;\tilde{u}_{T2}\;\tilde{u}_{N2}]^T&=&\dfrac{r_2}{\sqrt{\mu p_2}}\, {\mathbf{u}}_2.
\end{array}
\end{equation}
Differentiating both sides of \eqref{abmap} with respect to time, substituting the Gauss' variational equations (GVEs \cite{battin1999introduction}) for ${i_j}$, ${\Omega_j}$, taking into account \eqref{tildeu}, and solving the resulting expression for $\dot{\alpha_1}$, $\dot{\alpha_2}$ and $\dot{\gamma}$ gives
\begin{equation}\label{dotalpha}
\dot{\alpha_1}=\frac{\sin\theta_2}{\sin\gamma}\,\tilde{u}_{N2}-[\sin \theta_1 \cot\gamma+\sin(\theta_1+\alpha_1)\cot i_1]\,\tilde{u}_{N1}
\end{equation}
\begin{equation}\label{dotbeta}
\dot{\alpha_2}=[\sin \theta_2 \cot\gamma-\sin(\theta_2+\alpha_2)\cot i_2]\,\tilde{u}_{N2}-\frac{\sin\theta_1}{\sin\gamma}\,\tilde{u}_{N1}
\end{equation}
\begin{equation}\label{dotgamma}
\dot{\gamma}=\cos\theta_2\,\tilde{u}_{N2}-\cos\theta_1\,\tilde{u}_{N1}.
\end{equation}
By using \eqref{releo}, \eqref{tildeu}-\eqref{dotbeta} and the GVEs for the arguments of latitude and of the periapsis, we get
\begin{equation}\label{dtt1}
\dot{\theta}_1=\frac{\sqrt{\mu p_1}}{r_1^2}+\sin \theta_1 \cot\gamma \,\tilde{u}_{N1}-\frac{\sin\theta_2}{\sin\gamma}\,\tilde{u}_{N2}
\end{equation}
\begin{equation}
\dot{\theta}_2=\frac{\sqrt{\mu p_2}}{r_2^2}-\sin \theta_2 \cot\gamma\,\tilde{u}_{N2}+\frac{\sin\theta_1}{\sin\gamma}\,\tilde{u}_{N1}
\end{equation}
\begin{equation}
\dot{\lambda}_1=\frac{p_1\!+\!r_1}{r_1 e_1}\sin v_1\,\tilde{u}_{T1}-\frac{p_1}{r_1 e_1}\cos v_1\,\tilde{u}_{R1}+\sin \theta_1 \cot\gamma \,\tilde{u}_{N1}-\frac{\sin\theta_2}{\sin\gamma}\,\tilde{u}_{N2}
\end{equation}
\begin{equation}\label{dll2}
\dot{\lambda}_2=\frac{p_2\!+\!r_2}{r_2 e_2}\sin v_2\,\tilde{u}_{T2}-\frac{p_2}{r_2 e_2}\cos v_2\,\tilde{u}_{R2}-\sin \theta_2 \cot\gamma\,\tilde{u}_{N2}+\frac{\sin\theta_1}{\sin\gamma}\,\tilde{u}_{N1}.
\end{equation}
The expressions \eqref{dotgamma}-\eqref{dll2} provide a convenient set of variational equations for the relative inclination angle and the four nodal elements.

By differentiating \eqref{mypar} with respect to time and using \eqref{tildeu},\eqref{dotgamma}-\eqref{dll2} together with the GVEs for $p_j$ and $e_j$, one can express the perturbed relative motion dynamics as
\begin{equation}\label{pertreldyn}
\dot{\textbf{\oe}}=\mathbf{f}(\textbf{\oe};\boldsymbol{\eta})+\mathbf{G}_2(\textbf{\oe};\boldsymbol{\eta}) \,{\mathbf{u}}_2-\mathbf{G}_1(\textbf{\oe};\boldsymbol{\eta})\, {\mathbf{u}}_1,
\end{equation}
where $\mathbf{f}(\textbf{\oe};\boldsymbol{\eta})$ is specified by \eqref{mypardyn2}-\eqref{dotv1},
\begin{equation}\label{g2inp}
\mathbf{G}_2(\textbf{\oe};\boldsymbol{\eta})=
\dfrac{r_2}{\sqrt{\mu p_2}}
\left[
\begin{array}{ccc}
0& 0 &\delta h_\theta\\[2mm]
0& 2(1+\delta p) &0 \\[2mm]
\dfrac{p_2}{r_2}\sin \delta\theta& \dfrac{2p_2}{r_2}\cos \delta\theta+e_{\theta}\sin\delta\theta &(\delta\xi_y+e_1\sin v_1)\, \delta h_\theta \\[2mm]
\dfrac{p_2}{r_2}\cos \delta\theta& -\dfrac{2p_2}{r_2}\sin \delta\theta+e_{\theta}\cos\delta\theta  &-(\delta\xi_x+e_1\cos v_1)\,\delta h_\theta \\[2mm]
0& 0& \dfrac{1+\delta h_x^2+\delta h_y^2}{2}\cos \delta\theta \\[2mm]
0& 0 &-\dfrac{1+\delta h_x^2+\delta h_y^2}{2}\sin \delta\theta
\end{array}\right],
\end{equation}
\begin{equation}\label{g1inp}
\mathbf{G}_1(\textbf{\oe};\boldsymbol{\eta})=
\dfrac{r_1}{\sqrt{\mu p_1}}
\left[
\begin{array}{ccc}
0& 0 &-\delta h_y\\[2mm]
0& 2(1+\delta p) &0\\[2mm]
0& 2{p_1}/{r_1} &-(\delta\xi_y+e_1\sin v_1)\,\delta h_y\\[2mm]
{p_1}/{r_1}& e_1\sin v_1 &(\delta\xi_x+e_1\cos v_1)\,\delta h_y\\[2mm]
0& 0& (1+\delta h_x^2-\delta h_y^2)/2 \\[2mm]
0& 0 &\delta h_x \delta h_y
\end{array}\right], \vspace{3mm}
\end{equation}
$r_1$, $r_2$, $p_2$ can be expressed in terms of $\textbf{\oe}$ and $\boldsymbol{\eta}$ via \eqref{mypar},\eqref{radius1}-\eqref{radius2}, and
\begin{equation}
\begin{array}{c}
\delta h_\theta=\delta h_x\sin\delta\theta+\delta h_y\cos\delta\theta,\\
e_{\theta}=(\delta\xi_x+e_1\cos v_1)\sin\delta\theta+ (\delta\xi_y+e_1\sin v_1)\cos\delta\theta.
\end{array}
\end{equation}
The perturbed dynamics of the parameter vector \eqref{etavec} are obtained from the GVEs for $v_1$, $p_1$ and $e_1$ as
\begin{equation}\label{pvdyn}
\dot{\boldsymbol{\eta}}=\mathbf{f}_\eta(\boldsymbol{\eta}) +\mathbf{G}_\eta(\boldsymbol{\eta})\,\mathbf{u}_1,
\end{equation}
where $\mathbf{f}_\eta(\boldsymbol{\eta})$ is given by \eqref{uvdyn2}, and
\begin{equation}\label{gvinp}
\mathbf{G}_\eta(\boldsymbol{\eta})=
\dfrac{r_1}{\sqrt{\mu p_1}}
\left[
\begin{array}{ccc}
0& 2p_1 &0\\[2mm]
0& 2{p_1}/{r_1} & 0\\[2mm]
{p_1}/{r_1}& e_1\sin v_1&0
\end{array}\right].
\end{equation}
The 9-dimensional nonlinear model \eqref{pertreldyn}-\eqref{gvinp} can be used to account for both conservative and nonconservative perturbations affecting the two satellites (for a detailed description of such contributions see, e.g., \cite{canuto2018spacecraft}). The model is valid for arbitrary elliptical orbits, except for purely (relatively) retrograde ones.

\subsection{Model Validation}\label{sima}

\begin{figure}[!t]
\centering
\includegraphics[width=0.55\textwidth]{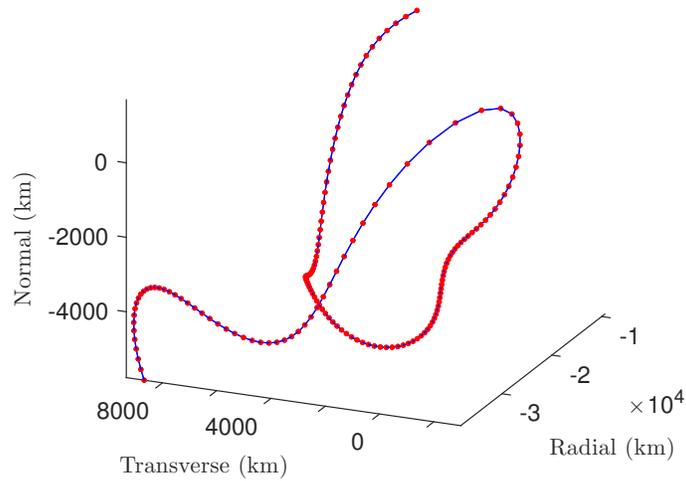}   %submit
\caption{The relative position trajectories obtained with the the proposed model (solid) and with Cowell's formulation (dotted) show an exact matching.}\label{valid}
\end{figure}

Model \eqref{pertreldyn}-\eqref{gvinp} has been validated against the standard Cowell's formulation in a Matlab simulation environment. Simulation results are reported for two sample orbits defined in terms of the initial orbital elements
\begin{equation*}
\begin{array}{l l l}
\{a_1,e_1,i_1,\Omega_1,\omega_1,v_1\}&=&\{8.9 \cdot 10^3\, \text{km},0.5,10^\circ,20^\circ,0^\circ,30^\circ \},
\\
\{a_2,e_2,i_2,\Omega_2,\omega_2,v_2\}&=&\{6.8 \cdot 10^3\, \text{km},0.1,40^\circ,90^\circ,30^\circ,70^\circ\}.
\end{array}
\end{equation*}
The initial conditions for the propagation of \eqref{pertreldyn}-\eqref{g1inp} are obtained by using \eqref{releo}-\eqref{abmap} and \eqref{mypar}.
%In particular, the following line of code is used to extract the angles $-\alpha_2$, $\gamma$, and $\alpha_1$ from \eqref{abmap}:\\
%\texttt{[mbet,gam,alph]=dcm2angle(angle2dcm(i2,W2-W1,-i1,'XZX'),'ZXZ')}.\\
The components of the input vectors $\mathbf{u}_1$ and $\mathbf{u}_2$ are specified as sinusoidal signals with different frequencies and phases, and amplitude equal to $1$ m/s$^2$. The output of \eqref{pertreldyn}-\eqref{gvinp} is transformed into local relative position coordinates by using \eqref{orbmotion}-\eqref{posquotient}. The resulting trajectory is depicted together with that obtained from Cowell's formulation in Fig.~\ref{valid}, on a time interval of $10^4$~s. As expected, there is an exact matching between the two solutions.

\section{Collision Detection}\label{codect}
As outlined in Section \ref{RMD}, the parameters \eqref{errorstate}-\eqref{etavec} allow one to fully characterize the relative motion geometry.
In this section, we will exploit this feature to derive a simple and computationally efficient collision detection technique. In order for two satellite to collide, the following conditions (denoted by C1 and C2) must be verified: (C1) their orbits intersect and (C2) the satellites pass through the intersection at the same time. For the purpose of collision detection, it is of interest to determine in advance whether these conditions could be met. While C1 is fully specified by the shape and the relative orientation of the two satellite orbits (i.e., by the Keplerian invariants), evaluating C2 generally requires a time-consuming numerical propagation of the relative motion. A possible way to mitigate this issue is to first determine if C1 is satisfied, and then check C2 only if needed. In collision avoidance applications, it is convenient to falsify C1, which alleviates the need for evaluating C2. The orbits for which C1 is falsified are commonly referred to as \emph{passively safe}. For close formations in near circular orbits, it has been shown in \cite{d2006proximity} that passive safety is strongly related to the phase separation between the relative eccentricity and inclination vectors. Hereafter, this concept is specialised for collision detection in elliptical orbits.

By using \eqref{orbmotion} and the fact that $b_1^2+b_2^2+b_3^2=1$, the inter-satellite distance can be expressed as
\begin{equation}\label{relrad}
\|\delta \mathbf{r}\|=r_1\sqrt{1+q^2-2 q b_1}.
\end{equation}
In the event of a collision, $\|\delta \mathbf{r}\|=0$. By enforcing this constraint in \eqref{relrad}, one gets
\begin{equation}\label{cevent}
1+q^2-2 q b_1=0.
\end{equation}
Equation \eqref{cevent} corresponds to both conditions C1 and C2. In order to isolate C1, we will exploit the key geometrical features of the proposed parametrization. Coplanar and noncoplanar orbits are treated separately.

In the coplanar case, one has $\delta h_x=\delta h_y=0$ and the collision condition $\delta\theta=0$ (see \eqref{mypar} and Fig.~\ref{relel2}). Then, according to \eqref{posquotient}, $b_1=1$  and \eqref{cevent} becomes
\begin{equation}\label{ceventc}
q=1,
\end{equation}
which is expected, since $q=r_2/r_1$. By using $\delta\theta=0$ in \eqref{posquotient} and substituting the expression for $q$ in \eqref{ceventc}, one gets
\begin{equation}\label{ceventcs}
\delta p(1+e_1 \cos v_1)=\delta\xi_x.
\end{equation}
For $\delta\theta=0$, the parameter $\delta\xi_x$ takes on the form
\begin{equation}\label{deltasig0}
\delta \xi_x=e_2\cos (v_1-\delta\lambda)-e_1 \cos v_1,
\end{equation}
where the Keplerian invariants $e_2$ and $\delta \lambda=\lambda_2-\lambda_1$ can be expressed in terms of \eqref{errorstate}-\eqref{etavec} as
\begin{equation}\label{ecc2par}
e_2=\sqrt{\delta \xi^2+e_1^2+2 e_1(\delta \xi_x\cos v_1+\delta \xi_y\sin v_1)}
\end{equation}
\begin{equation}\label{dlambda}
\delta \lambda=\text{atan}_2\left(\delta \xi_x\sin v_1-\delta \xi_y\cos v_1,\;\delta \xi_x\cos v_1 +\delta \xi_y\sin v_1  +e_1\right).
\end{equation}
Substituting \eqref{deltasig0} into \eqref{ceventcs} and simplifying the resulting expression gives
\begin{equation}\label{ceventcse}
\delta p+\ \delta\varrho\cos \left(v_1+\phi_c \right)=0
\end{equation}
where
\begin{equation}\label{vrhoeq}
\delta\varrho=\sqrt{(1+\delta p)^2e_1^2+e_2^2-(1+\delta p)e_1 e_2\cos(\delta\lambda)}
\end{equation}
\begin{equation}
\phi_c=\text{atan}_2\Big(e_2\sin(\delta \lambda),\, (1+\delta p)e_1-e_2\cos (\delta \lambda)\Big).
\end{equation}
In order for C1 to hold, there must exist a $v_1$ such that \eqref{ceventcse} is satisfied. This occurs if and only if $\delta p^2 -\delta\varrho^2 \leq 0$.
This condition reduces to $\delta p^2 - \delta\xi^2 \leq 0$ when $e_1=0$ (see \eqref{ecc2par} and \eqref{vrhoeq}).

For noncoplanar orbits, one has the collision conditions $\theta_1=\theta_2=0$ (see Fig.~\ref{relel2}) and $\theta_1=\theta_2=\pi$ (at the opposite relative line of nodes crossing), which again imply \eqref{ceventc}. By observing, from \eqref{releo} and \eqref{eivrel}, that $v_1=-\lambda_1$ and $\delta\xi_x=\delta e_x$ at $\theta_1=0$, and using these identities in \eqref{ceventcs}, one gets the condition
$\delta p(1+e_1 \cos \lambda_1)-\delta e_x=0$. Similarly, for $\theta_1=\pi$, one gets $\delta p(1-e_1 \cos \lambda_1)+\delta e_x=0$, where $\delta e_x=\delta \xi \cos(\delta\phi)$ according to \eqref{eivecex}-\eqref{eccphase}. Summarizing, condition C1 can be expressed as
\begin{equation}\label{ellcoll}
\text{C1}:\;\,
\left\{
\begin{array}{c c l}
\delta p^2-\;\delta\varrho^2 \,\leq 0&\quad& \text{if}\;\; \delta h= 0\\
\delta p\, (1+e_1 \cos \lambda_1)-\delta \xi \cos(\delta\phi)=0 &\quad& \text{if}\;\; \delta h\neq 0 \;\text{(ascending)}\\
\delta p\, (1-e_1 \cos \lambda_1)+\delta \xi \cos(\delta\phi)=0 &\quad& \text{if}\;\; \delta h\neq 0 \;\text{(descending)}
\end{array}
\right.
\end{equation}
Condition \eqref{ellcoll} provides a quick test to rule out potential collisions, possibly occurring at later times. For circular reference orbits, \eqref{ellcoll} is fully specified by the invariants $\{\delta p,\,\delta \xi,\, \delta h ,\,\delta\phi\}$ which, in turn, depend only on the relative state vector \eqref{errorstate}. For noncoplanar orbits with equal semiparameter ($\delta h\neq 0$, $\delta p=0$), the collision safety margin turns out to be proportional to the magnitude of the term  $\delta \xi \cos(\delta\phi)$, and thus it is maximized for parallel eccentricity and inclination vector configurations ($\delta\phi=0,\,\pi$). Conversely, two orbits such that $\delta h\neq 0$ and $\delta p=0$ are unsafe if $|\delta\phi|= \pi/2$.

In order to check if a collision actually occurs within a certain time $t_f$ (condition C2), one can propagate the relative motion to determine if $\|\delta \mathbf{r}(t)\|=0$ (within a given tolerance) for $t\in[t_0,\; t_f]$, by using the motion dynamics derived in Section \ref{rmotmod}.

\section{Relative Navigation Scheme}\label{aanavigation}
 The following angles-only relative navigation problem is considered, as a benchmark for the proposed modeling approach. Let satellite 1 be able to measure the azimut and elevation angles of vector $\delta\mathbf{r}$ in the RTN$_1$ frame, as well as the apparent angular size $\beta$ of satellite 2. For simplicity, satellite 2 is modeled as a sphere with diameter $d$. The measurement model for $\varphi_{\text{az}}$, $\varphi_{\text{el}}$ and $\beta$ is given by \cite{woffinden07}
\begin{equation}\label{measeq}
\begin{array}{l l l}
\varphi_{\text{az}}&=&\text{atan}_2\left(\delta{r}_{T},\delta{r}_{R} \right)+w_\text{az}\\
\varphi_{\text{el}}&=&\text{asin}\left(\delta {r}_{N}/\| \delta \mathbf{r} \|\right)+w_\text{el}\\
\beta&=&d/\| \delta \mathbf{r}\|+w_\beta,
\end{array}
\end{equation}
where $\delta\mathbf{r}=[\delta{r}_{R}\;\delta{r}_{T}\; \delta{r}_{N}]^T$ and $w_\text{az}$, $w_\text{el}$, $w_\beta$ are discrete-time white noise processes modeling the measurement error. The measurement model can be expressed in terms of \eqref{errorstate}-\eqref{etavec} as
\begin{equation}\label{obsmod}
\check{\mathbf{y}}=\mathbf{y}({\textbf{\oe}};{\boldsymbol{\eta}})+\mathbf{w},
\end{equation}
where $\check{\mathbf{y}}=[\varphi_{\text{az}}\;\, \varphi_{\text{el}}\;\,\beta]^T$, $\mathbf{w}=[w_\text{az}\;\,w_\text{el}\;\,w_\beta]^T$, and the
function $\mathbf{y}({\textbf{\oe}};{\boldsymbol{\eta}})$ is obtained from \eqref{radius1}, \eqref{orbmotion}-\eqref{posquotient} and \eqref{measeq}. Our objective is to estimate the relative state vector \eqref{errorstate} from a set of angular measurements of the form \eqref{measeq}. It is assumed that the parameter vector $\boldsymbol{\eta}$, the diameter $d$, and the gravitational parameter $\mu$ are known, and that both satellites are uncontrolled.

An extended Kalman filtering approach is adopted for relative navigation. The estimated state is denoted by $\hat{\textbf{\oe}}$. Environmental perturbations are treated as process disturbances. This approximation is reasonable as long as the relative motion is dictated mainly by spherical gravity. The disturbance-free dynamics of the filter estimates are then given by
\begin{equation}\label{filterdyn}
\hat{\dot{\textbf{\oe}}}=\mathbf{f}(\hat{\textbf{\oe}};{\boldsymbol{\eta}}).
\end{equation}
The EKF propagation and update steps are performed as per standard, see, e.g., \cite{ceccarelli2007spacecraft}. Notice that in the propagation step one can exploit the analytic solutions $\delta p(t)=\delta p(t_0)$ and \eqref{sigmasol}-\eqref{hsol}.

In the next section, the performance of the EKF scheme is evaluated on an asteroid flyby application.

\section{Asteroid Flyby Application}\label{simb}
On 10 July 2010, the ESA Rosetta spacecraft performed a close flyby of asteroid Lutetia during its journey to the target comet 67P/Churyumov-Gerasimenko. Rosetta passed the asteroid at a distance of 3170 km, with a relative velocity of 15 km/s. About one month prior to the encounter, an optical observation campaign was started to assist the asteroid orbit determination \cite{SCHULZ20122}.

\begin{figure}
\centering
\includegraphics[width=0.65\textwidth]{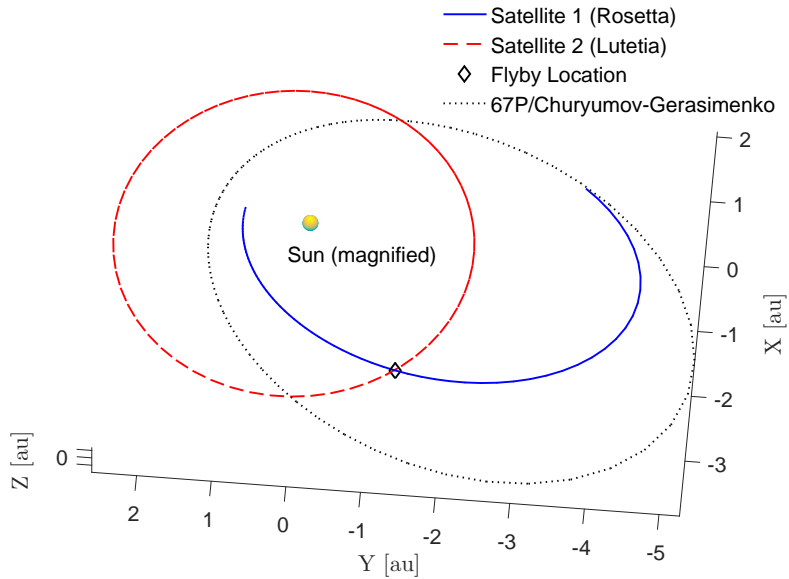}
\caption{Asteroid intercept problem geometry.}
\label{rosfig}
\end{figure}
In the following we consider a similar scenario and assume that the spacecraft (satellite 1) and the asteroid (satellite 2) will eventually collide, unless an evasive maneuver is performed. The geometry of the intercept problem is depicted in Fig.~\ref{rosfig}. This problem gives us the opportunity to highlight some of the salient features of the proposed modeling approach, including the applicability to elliptical orbits, and its suitability for collisions avoidance applications. The EKF scheme in Section \ref{aanavigation} is used to assess the performance of angles-only relative navigation for the considered problem. Condition C1 in \eqref{ellcoll} is evaluated on the EKF estimates to check whether the collision can be detected with sufficient notice.

\begin{figure}[t]
\centering
\includegraphics[width=0.8\textwidth]{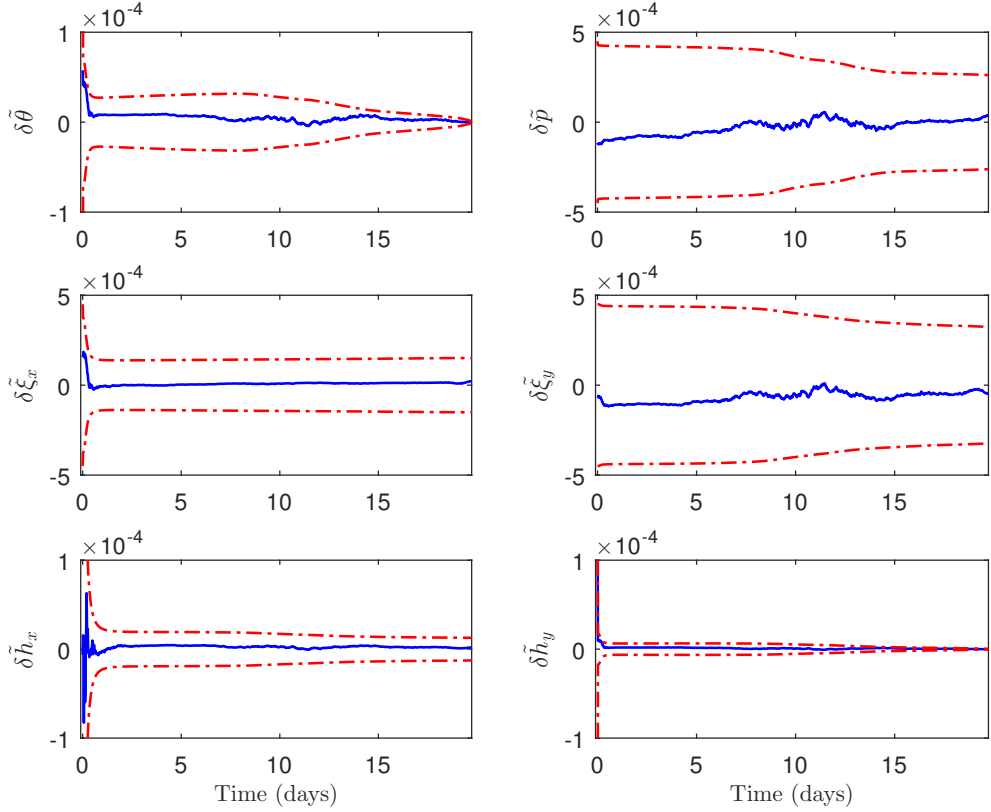}
\caption{Evolution of the EKF estimation error with $3\sigma$ confidence intervals.}
\label{ekfps}
\end{figure}
The mission is simulated on a time interval ranging from 20 days to 6 hours before impact. The resulting satellite trajectories pass through a small neighborhood of the flyby location in Fig.~\ref{rosfig}. The diameter $d$ and the standard deviation of the measurement noise in \eqref{measeq} are set to $d=90$ km and $\sigma_\text{az}=\sigma_\text{el}=\sigma_\beta=0.001$ deg. These values are consistent with the size of Lutetia and with the specifications of the optical instruments installed on-board Rosetta, respectively. Measurements are collected once every 5 s. The EKF is initialized by adding a random perturbation vector to the true initial state $\textbf{\oe}(t_0)$. The standard deviation of each element of the perturbation vector is set to $1.5\cdot10^{-4}$. The resulting initial relative position uncertainty is in the order of $10^5$ km, and it is much larger than the uncertainty affecting the asteroid ephemeris. The initial value of the $6 \times 6$ estimation error covariance matrix $\mathbf{P}$ is set accordingly. Hereafter, we will first present the results of a single simulation to show how the EKF estimates can be used for collision avoidance and then a Monte Carlo analysis of the filter performance will be reported.

\begin{figure}[t]
\centering
\includegraphics[width=0.45\textwidth]{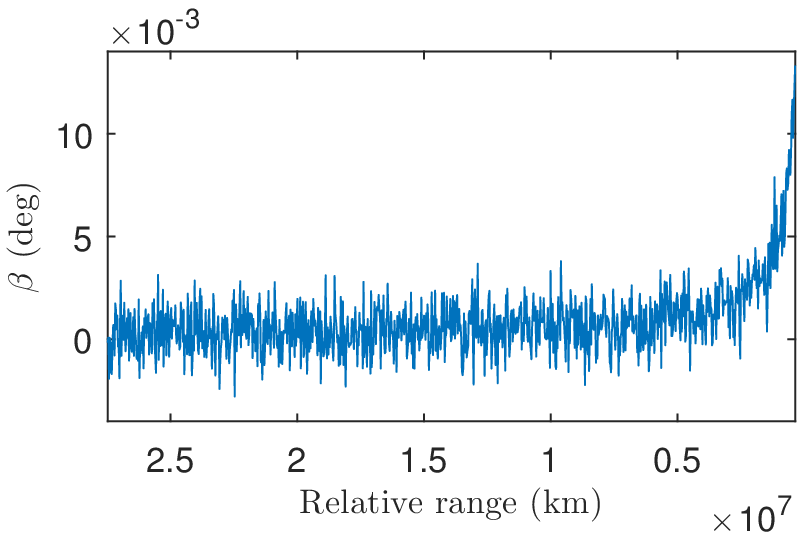}
\caption{Evolution of angular size measurements versus relative range: for relative range values greater than $5\cdot10^6$ km, the apparent angular size of satellite 2 is smaller than the pixel size, which is equal to the measurement noise standard deviation.}
\label{asmeas}
\end{figure}

The components of the estimation error vector $\tilde{\textbf{\oe}}=\hat{\textbf{\oe}}-{\textbf{\oe}}$ are shown in Fig.~\ref{ekfps} for a sample simulation, together with the corresponding $3\sigma$ confidence intervals (obtained from matrix $\mathbf{P}$). It can be seen that all estimates are consistent, although some relative states are estimated with less confidence than others. Notice that this problem is not related to the specific parametrization employed, but rather to the impossibility to extract range information from the noisy measurements \eqref{measeq} prior to the final approach phase (see Fig.~\ref{asmeas}), coupled with the relatively short period in which the asteroid can be observed compared to the orbital period. This affects the quality of state reconstruction from angular measurements, and results in observability issues for some of the relative states. In particular, the path followed by the two satellites within the observation window is almost rectilinear, which makes it difficult to estimate the relative states describing the orbit shape (semiparameter, eccentricity). However, this does not prevent the filter from accurately estimating the safety-critical parameters (relative range, collision safety margin).

\begin{figure}[!t]
\centering
\includegraphics[width=0.65\textwidth]{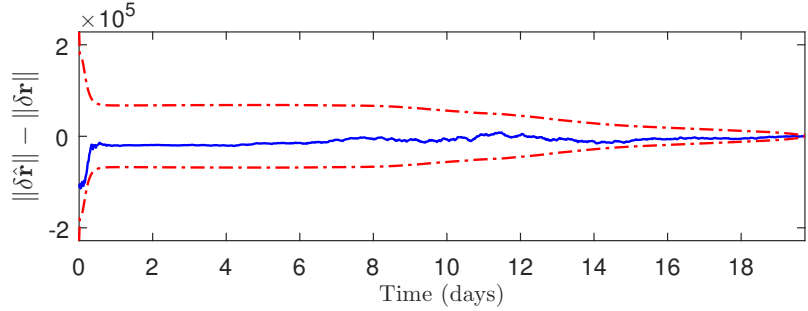}
\caption{Evolution of the relative range estimation error with $3\sigma$ confidence interval.}
\label{dmeas}
\end{figure}

The relative range estimation error $\|\delta \hat{\mathbf{r}}\|-\|\delta {\mathbf{r}}\|$ and its $3\sigma$ confidence interval are reported in Fig.~\ref{dmeas}. After the initial transient, they settle to a steady state condition in which basically no correction is performed, until the satellites are close enough and the angular size measurements become effective (see Fig.~\ref{asmeas}). The estimation error at the end of the simulation amounts to $273$ km. It is about 3 orders of magnitude smaller than both the initialization error and the final relative range between the two satellites. Given the limitations of angles-only measurements for the considered intercept problem, this is a fairly good figure.

\begin{figure}[!t]
\centering
\includegraphics[width=0.65\textwidth]{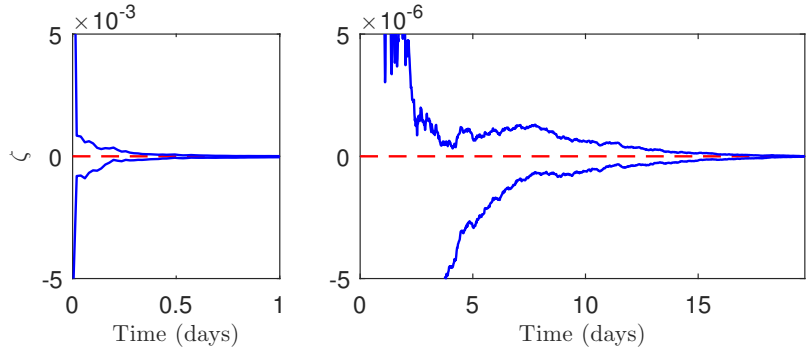}
\caption{Evolution of the $3\sigma$ confidence interval centered at the estimate $\zeta(\hat{\textbf{\oe}};\boldsymbol{\eta})$: initial transient (left) and convergence toward the steady state (right); notice the different $y$-axis scales.}
\label{coller}
\end{figure}
Concerning the ability to detect collisions, the first and the third condition in \eqref{ellcoll} are not satisfied by construction (the two orbits are noncoplanar and the collision is set to occur at $\theta_1=\theta_2=0$). The second condition in \eqref{ellcoll} is rewritten as $\zeta(\textbf{\oe};\boldsymbol{\eta})=0$, where
\begin{equation}\label{collpar}
\begin{array}{l l l}
\zeta(\textbf{\oe};\boldsymbol{\eta})&=&\delta p(1+e_1 \cos \lambda_1)-\delta \xi \cos (\delta\phi)\\[2mm]
&=&\delta p-\dfrac{\delta h_x (\delta \xi_x-\delta p\, e_1\!\cos v_1)+\delta h_y(\delta \xi_y-\delta p\, e_1\!\sin v_1)}{(\delta h_x^2+\delta h_y^2)^{1/2}}
\end{array}
\end{equation}
can be thought of as the collision safety margin. The $3\sigma$ confidence interval centered at the estimate $\zeta(\hat{\textbf{\oe}};\boldsymbol{\eta})$ is adopted to characterize the parameter uncertainty region. It is reported in Fig.~\ref{coller}, and contains the value $\zeta=0$ over the entire simulation interval. Hence, a potential collision is successfully detected. The uncertainty region shrinks down to a very small neighborhood of the origin after a short transient, which indicates that the estimate $\zeta(\hat{\textbf{\oe}};\boldsymbol{\eta})$ is unaffected by the above-mentioned observability issues. This is not surprising, since estimating the intersection between two almost-rectilinear orbital segments is considerably simpler than estimating the curvature of the orbits along which they lie.

\begin{figure}[t]
\centering
\includegraphics[width=0.45\textwidth]{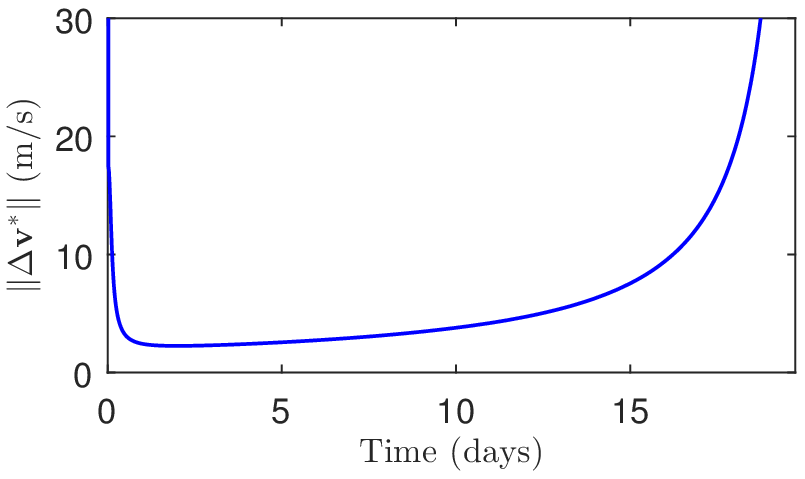}
\caption{Delta-v profile computed from \eqref{dvopt} along the trajectories of \eqref{filterdyn}.}
\label{dvav}
\end{figure}
The information provided by the EKF estimate $\zeta(\hat{\textbf{\oe}};\boldsymbol{\eta})$ allows one to plan a timely collision avoidance maneuver. The maneuver consists of correcting $\zeta$ by a suitable $\Delta\zeta$, so as to falsify condition C1. The following impulsive maneuvering strategy can be adopted to this purpose. Let $\mathbf{u}_2=\mathbf{0}$ and $\mathbf{u}_1=\Delta\mathbf{v}\, \delta(t_m)$, where $\Delta\mathbf{v}$ is an instantaneous velocity change applied to satellite 1, and $\delta(t_m)$ is a Dirac delta function centered at the impulse time $t_m$. Taking the time derivative of \eqref{collpar} $\frac{\partial{\zeta}(\textbf{\oe};\boldsymbol{\eta})}{\partial \boldsymbol{\eta}}\dot{\boldsymbol{\eta}} + \frac{\partial{\zeta}(\textbf{\oe};\boldsymbol{\eta})}{\partial \textbf{\oe}}\dot{ \textbf{\oe}}$, substituting the above-defined inputs in \eqref{pertreldyn} and \eqref{pvdyn}, and integrating the resulting expression at $t_m$ gives
\begin{equation}\label{dsig}
\Delta\zeta=\left(\frac{\partial\zeta(\textbf{\oe};\boldsymbol{\eta})}{\partial \boldsymbol{\eta}}\mathbf{G}_\eta(\boldsymbol{\eta})-\frac{\partial\zeta(\textbf{\oe};\boldsymbol{\eta})}{\partial \textbf{\oe}}\mathbf{G}_1(\textbf{\oe};\boldsymbol{\eta})\right)\Delta\mathbf{v}=\mathbf{g}^T(\textbf{\oe};\boldsymbol{\eta}) \Delta\mathbf{v},
\end{equation}
where $\Delta\zeta$ is the instantaneous change in $\zeta$ corresponding to $\Delta\mathbf{v}$, and all quantities are evaluated at $t_m$. The $\Delta\mathbf{v}$ required to satisfy \eqref{dsig} has minimum norm for
\begin{equation}\label{dvv}
\Delta\mathbf{v}=\Delta\mathbf{v}^\ast(\textbf{\oe};\boldsymbol{\eta})=\frac{\mathbf{g}(\textbf{\oe};\boldsymbol{\eta})}{\| \mathbf{g}(\textbf{\oe};\boldsymbol{\eta}) \|^2} \,\Delta\zeta.
\end{equation}
From \eqref{dvv}, it follows that
\begin{equation}\label{dvopt}
\|\Delta\mathbf{v}^\ast(\textbf{\oe};\boldsymbol{\eta})\|=\frac{|\Delta\zeta|}{\| \mathbf{g}(\textbf{\oe};\boldsymbol{\eta}) \|}.
\end{equation}
To assess the feasibility of the collision avoidance maneuver, the impulse magnitude $\|\Delta\mathbf{v}^\ast(\textbf{\oe};\boldsymbol{\eta})\|$ in \eqref{dvopt} has been evaluated at different time instants along the estimated trajectory. The correction term $\Delta\zeta$ is specified as the sum of the estimated  $3\sigma$ uncertainty on $\zeta(\hat{\textbf{\oe}};\boldsymbol{\eta})$ and a constant offset. The offset value is set to $10^{-4}$, which corresponds to a nominal displacement of approximately 3000 km at the closest approach point. From day $1$ (end of initial estimation transient) till day $7$, the resulting $\|\Delta\mathbf{v}^\ast(\hat{\textbf{\oe}};\boldsymbol{\eta})\|$ turns out to be smaller than 3 m/s, while it rapidly grows subsequently, see Fig.~\ref{dvav}. Such a velocity change translates into a negligible fuel expenditure and it is fully compatible with a mission of the considered type.

\begin{figure}[t]
\centering
\includegraphics[width=0.8\textwidth]{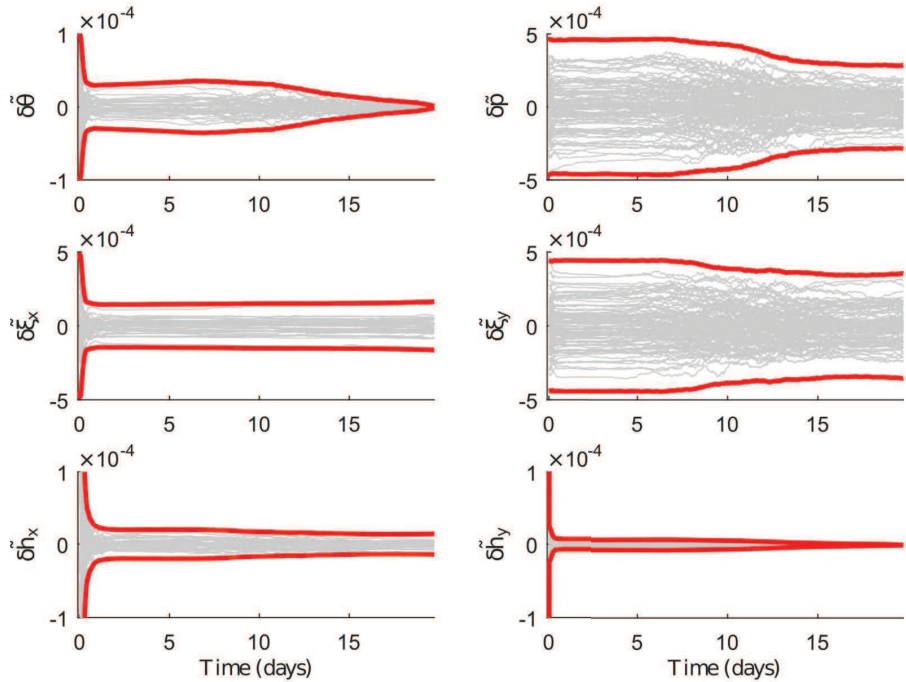}
\caption{Monte Carlo results for the EKF estimation error: true $3\sigma$ confidence intervals (thick lines) and trajectory samples  (thin lines).}
\label{ekfpsmc}
\end{figure}
\begin{figure}[t]
\centering
\includegraphics[width=0.65\textwidth]{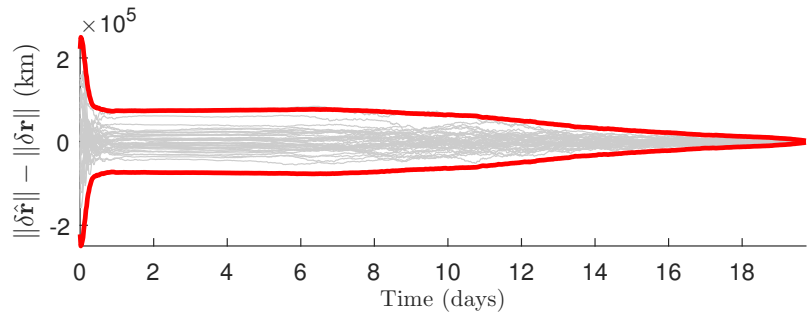}
\caption{Monte Carlo results for  the relative range estimation error: true $3\sigma$ confidence interval (thick lines) and trajectory samples  (thin lines).}
\label{dmeasmc}
\end{figure}
\begin{figure}[t]
\centering \psfrag{\varsigma}{$\zeta$}
\includegraphics[width=0.65\textwidth]{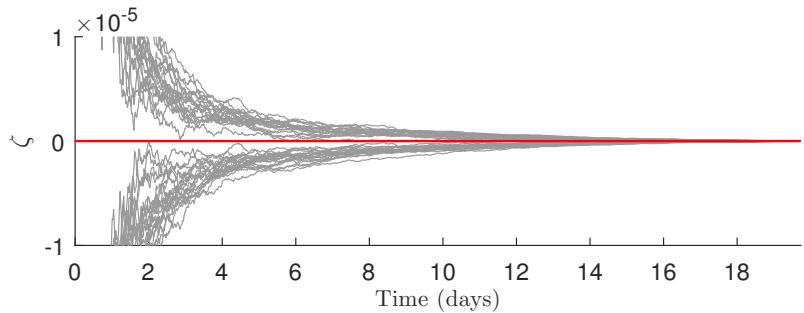}
\caption{Monte Carlo results for the confidence intervals centered at the estimate $\zeta(\hat{\textbf{\oe}};\boldsymbol{\eta})$ (thin lines).}
\label{collermc}
\end{figure}
The statistical results from 200 Monte Carlo runs of the flyby simulation are reported in Figs.~\ref{ekfpsmc}-\ref{collermc} for the navigation, relative range and collision detection errors. The $3\sigma$ confidence intervals computed from the true covariance, which are reported in Figs.~\ref{ekfpsmc}-\ref{dmeasmc}, agree nicely with that estimated by the EKF in Figs.~\ref{ekfps} and~\ref{dmeas}. The true standard deviation of the relative range estimation error at the end of the simulation is of 638 km. According to Fig.~\ref{collermc}, the potential collision is successfully detected in all realizations. From these results, it can be concluded that the EKF scheme provides a reliable estimate of the navigation states, thus allowing to safely avoid the collision.

\section{Conclusions}\label{conclu}
A new parametrization of the relative motion problem has been presented, which combines the benefits of relative orbital elements and local translational states. Additionally, it provides a simple characterization of colliding orbits. The forward nonlinear mappings from classical orbital elements to the proposed parameters, and from these parameters to local translational states, have been described. The Keplerian and the perturbed parameter dynamics have been derived. No simplifying approximations are made in the derivation and the resulting model is valid for arbitrary elliptical orbits, except for purely (relatively) retrograde ones. An angles-only navigation filter and a collision avoidance scheme have been proposed in order to validate the model. These have been tested on an asteroid flyby application. Simulation results indicate that a collision can be detected
early on in the estimation process, which allows one to issue a timely and reasonably small-sized evasive maneuver.

\bibliographystyle{aiaa-doi}
\bibliography{biblio}

\begin{thebibliography}{10}
\newcommand{\enquote}[1]{``#1''}

\bibitem{clohessy1960terminal}
Clohessy, W.~H. and Wiltshire, R.~S., \enquote{Terminal Guidance System for
  Satellite Rendezvous,} {\em Journal of Aerospace Sciences\/}, Vol.~27, No.~9,
  1960, pp.~653--658.
\newblock \doi{10.2514/8.8704}.

\bibitem{lawden1963optimal}
Lawden, D.~F., {\em Optimal Trajectories for Space Navigation\/}, Vol.~3,
  Butterworths, London, UK, 1963, pp.\ 79--86.

\bibitem{hempel1965rendezvous}
Hempel, P. and Tschauner, J., \enquote{Rendezvous with a Target in an
  Elliptical Orbit,} {\em Astronautica Acta\/}, Vol.~11, No.~2, 1965,
  pp.~104--109.

\bibitem{mattingly2011}
Mattingly, R. and May, L., \enquote{Mars Sample Return as a Campaign,} {\em
  2011 Aerospace Conference\/}, 2011, pp. 1--13.
\newblock \doi{10.1109/AERO.2011.5747287}.

\bibitem{landgraf2013}
Landgraf, M. and Mestreau-Garreau, A., \enquote{Formation Flying and Mission
  Design for Proba-3,} {\em Acta Astronautica\/}, Vol.~82, No.~1, 2013,
  pp.~137--145.
\newblock \doi{10.1016/j.actaastro.2012.03.028}.

\bibitem{Glassmeier2007}
Glassmeier, K.-H., Boehnhardt, H., Koschny, D., K{\"u}hrt, E., and Richter, I.,
  \enquote{The Rosetta Mission: Flying Towards the Origin of the Solar System,}
  {\em Space Science Reviews\/}, Vol.~128, No.~1, 2007, pp.~1--21.
\newblock \doi{10.1007/s11214-006-9140-8}.

\bibitem{TSUDA2013356}
Tsuda, Y., Yoshikawa, M., Abe, M., Minamino, H., and Nakazawa, S.,
  \enquote{System Design of the Hayabusa 2 - Asteroid Sample Return Mission to
  1999 JU3,} {\em Acta Astronautica\/}, Vol.~91, 2013, pp.~356 -- 362.
\newblock \doi{10.1016/j.actaastro.2013.06.028}.

\bibitem{Lauretta2017}
Lauretta, D.~S., Balram-Knutson, S.~S., Beshore, E., Boynton, W.~V.,
  Drouet~d'Aubigny, C., et~al., \enquote{OSIRIS-REx: Sample Return from
  Asteroid (101955) Bennu,} {\em Space Science Reviews\/}, Vol.~212, No.~1, Oct
  2017, pp.~925--984.
\newblock \doi{10.1007/s11214-017-0405-1}.

\bibitem{alfriend05}
Alfriend, K.~T. and Yan, H., \enquote{Evaluation and Comparison of Relative
  Motion Theories,} {\em Journal of Guidance, Control, and Dynamics\/},
  Vol.~28, No.~2, 2005, pp.~254--261.
\newblock \doi{10.2514/1.6691}.

\bibitem{sullivan2017comprehensive}
Sullivan, J., Grimberg, S., and D’Amico, S., \enquote{Comprehensive Survey
  and Assessment of Spacecraft Relative Motion Dynamics Models,} {\em Journal
  of Guidance, Control, and Dynamics\/}, Vol.~40, No.~8, 2017, pp.~1837--1859.
\newblock \doi{10.2514/1.G002309}.

\bibitem{Weiss12}
{Weiss}, A., {Kolmanovsky}, I., {Baldwin}, M., and {Erwin}, R.~S.,
  \enquote{Model Predictive Control of Three Dimensional Spacecraft Relative
  Motion,} {\em American Control Conference\/}, 2012, pp. 173--178.
\newblock \doi{10.1109/ACC.2012.6314862}.

\bibitem{quadrelli15}
Quadrelli, M.~B., Wood, L.~J., Riedel, J.~E., McHenry, M.~C., Aung, M.,
  Cangahuala, L.~A., Volpe, R.~A., Beauchamp, P.~M., and Cutts, J.~A.,
  \enquote{Guidance, Navigation, and Control Technology Assessment for Future
  Planetary Science Missions,} {\em Journal of Guidance, Control, and
  Dynamics\/}, Vol.~38, No.~7, 2015, pp.~1165--1186.
\newblock \doi{10.2514/1.G000525}.

\bibitem{dicairano18}
{Di Cairano}, S. and Kolmanovsky, I.~V., \enquote{Real-Time Optimization and
  Model Predictive Control for Aerospace and Automotive Applications,} {\em
  American Control Conference\/}, 2018, pp. 2392--2409.
\newblock \doi{10.23919/ACC.2018.8431585}.

\bibitem{ManciniCSL20}
{Mancini}, M., {Bloise}, N., {Capello}, E., and {Punta}, E., \enquote{Sliding
  Mode Control Techniques and Artificial Potential Field for Dynamic Collision
  Avoidance in Rendezvous Maneuvers,} {\em IEEE Control Systems Letters\/},
  Vol.~4, No.~2, 2020, pp.~313--318.
\newblock \doi{10.1109/LCSYS.2019.2926053}.

\bibitem{schaubrelative}
Schaub, H., \enquote{Relative Orbit Geometry Through Classical Orbit Element
  Differences,} {\em Journal of Guidance, Control, and Dynamics\/}, Vol.~27,
  No.~5, 2004, pp.~839--848.
\newblock \doi{10.2514/1.12595}.

\bibitem{gim2005satellite}
Gim, D.-W. and Alfriend, K.~T., \enquote{Satellite Relative Motion Using
  Differential Equinoctial Elements,} {\em Celestial Mechanics and Dynamical
  Astronomy\/}, Vol.~92, No.~4, 2005, pp.~295--336.
\newblock \doi{10.1007/s10569-004-1799-0}.

\bibitem{montenbruck2006}
Montenbruck, O., Kirschner, M., D'Amico, S., and Bettadpur, S.,
  \enquote{E/I-vector separation for safe switching of the GRACE formation,}
  {\em Aerospace Science and Technology\/}, Vol.~10, No.~7, 2006, pp.~628--635.
\newblock \doi{10.1016/j.ast.2006.04.001}.

\bibitem{d2006proximity}
D'Amico, S. and Montenbruck, O., \enquote{Proximity Operations of
  Formation-Flying Spacecraft Using an Eccentricity/Inclination Vector
  Separation,} {\em Journal of Guidance Control and Dynamics\/}, Vol.~29,
  No.~3, 2006, pp.~554--563.
\newblock \doi{10.2514/1.15114}.

\bibitem{d2010autonomous}
D'Amico, S., {\em Autonomous Formation Flying in Low Earth Orbit\/}, Ph.D.
  thesis, Technical University of Delft, The Netherlands, 2010.

\bibitem{vadali2002analytical}
Vadali, S.~R., \enquote{An Analytical Solution for Relative Motion of
  Satellites,} {\em Proceedings of the Fifth International Conference on
  Dynamics and Control of Structures and Systems in Space\/}, 2002, pp.
  309--316.

\bibitem{gurfil03}
Gurfil, P., \enquote{Control-Theoretic Analysis of Low-Thrust Orbital Transfer
  Using Orbital Elements,} {\em Journal of Guidance, Control, and Dynamics\/},
  Vol.~26, No.~6, 2003, pp.~979 -- 983.
\newblock \doi{10.2514/2.6926}.

\bibitem{hartley2012695}
Hartley, E.~N., Trodden, P.~A., Richards, A.~G., and Maciejowski, J.~M.,
  \enquote{Model Predictive Control System Design and Implementation for
  Spacecraft Rendezvous,} {\em Control Engineering Practice\/}, Vol.~20, No.~7,
  2012, pp.~695 -- 713.
\newblock \doi{10.1016/j.conengprac.2012.03.009}.

\bibitem{leomanni2017class}
Leomanni, M., Bianchini, G., Garulli, A., and Giannitrapani, A., \enquote{A
  Class of Globally Stabilizing Feedback Controllers for the Orbital Rendezvous
  Problem,} {\em International Journal of Robust and Nonlinear Control\/},
  Vol.~27, No.~18, 2017, pp.~4607--4621.
\newblock \doi{10.1002/rnc.3817}.

\bibitem{leomanni2018state}
Leomanni, M., Bianchini, G., Garulli, A., and Giannitrapani, A., \enquote{State
  Feedback Control in Equinoctial Variables for Orbit Phasing Applications,}
  {\em Journal of Guidance, Control, and Dynamics\/}, Vol.~41, No.~8, 2018,
  pp.~1815--1822.
\newblock \doi{10.2514/1.G003402}.

\bibitem{Gaias14}
Gaias, G., D’Amico, S., and Ardaens, J.-S., \enquote{Angles-Only Navigation
  to a Noncooperative Satellite Using Relative Orbital Elements,} {\em Journal
  of Guidance, Control, and Dynamics\/}, Vol.~37, No.~2, 2014, pp.~439--451.
\newblock \doi{10.2514/1.61494}.

\bibitem{sullivan17}
Sullivan, J. and D’Amico, S., \enquote{Nonlinear Kalman Filtering for
  Improved Angles-Only Navigation Using Relative Orbital Elements,} {\em
  Journal of Guidance, Control, and Dynamics\/}, Vol.~40, No.~9, 2017,
  pp.~2183--2200.
\newblock \doi{10.2514/1.G002719}.

\bibitem{kasdin2005canonical}
Kasdin, N.~J., Gurfil, P., and Kolemen, E., \enquote{Canonical Modelling of
  Relative Spacecraft Motion via Epicyclic Orbital Elements,} {\em Celestial
  Mechanics and Dynamical Astronomy\/}, Vol.~92, No.~4, 2005, pp.~337--370.
\newblock \doi{10.1007/s10569-004-6441-7}.

\bibitem{vallado2001fundamentals}
Vallado, D.~A., {\em Fundamentals of Astrodynamics and Applications\/},
  Springer-Verlag New York, 2nd ed., 2007, pp.\ 161-165.

\bibitem{gurfilmanifold}
Gurfil, P. and Kholshevnikov, K.~V., \enquote{Manifolds and Metrics in the
  Relative Spacecraft Motion Problem,} {\em Journal of Guidance, Control, and
  Dynamics\/}, Vol.~29, No.~4, 2006, pp.~1004--1010.
\newblock \doi{10.2514/1.15531}.

\bibitem{alfriend2009spacecraft}
Alfriend, K., Vadali, S.~R., Gurfil, P., How, J., and Breger, L., {\em
  Spacecraft Formation Flying: Dynamics, Control and Navigation\/}, Vol.~2,
  Butterworth Heinemann, Burlington, MA, 2009, pp.\ 147--150.

\bibitem{Kim07}
Kim, S.-G., Crassidis, J.~L., Cheng, Y., Fosbury, A.~M., and Junkins, J.~L.,
  \enquote{Kalman Filtering for Relative Spacecraft Attitude and Position
  Estimation,} {\em Journal of Guidance, Control, and Dynamics\/}, Vol.~30,
  No.~1, 2007, pp.~133--143.
\newblock \doi{10.2514/1.22377}.

\bibitem{woffinden07}
Woffinden, D.~C. and Geller, D.~K., \enquote{Relative Angles-Only Navigation
  and Pose Estimation For Autonomous Orbital Rendezvous,} {\em Journal of
  Guidance, Control, and Dynamics\/}, Vol.~30, No.~5, 2007, pp.~1455--1469.
\newblock \doi{10.2514/1.28216}.

\bibitem{battin1999introduction}
Battin, R.~H., {\em An Introduction to the Mathematics and Methods of
  Astrodynamics\/}, American Institute of Aeronautics and Astronautics, Inc.,
  Reston, Virginia, 1999, pp.\ 476-516.

\bibitem{canuto2018spacecraft}
Canuto, E., Novara, C., Carlucci, D., Montenegro, C.~P., and Massotti, L., {\em
  Spacecraft Dynamics and Control: The Embedded Model Control Approach\/},
  Butterworth-Heinemann, The Boulevard, Langford Lane, Kidlington, Oxford OX5
  1GB, United Kingdom, 2018, pp.\ 129-236.

\bibitem{ceccarelli2007spacecraft}
Ceccarelli, N., Garulli, A., Giannitrapani, A., Leomanni, M., and Scortecci,
  F., \enquote{Spacecraft Localization via Angle Measurements for Autonomous
  Navigation in Deep Space Missions,} {\em IFAC Proceedings Volumes\/},
  Vol.~40, No.~7, 2007, pp.~551 -- 556.
\newblock \doi{10.3182/20070625-5-FR-2916.00094}.

\bibitem{SCHULZ20122}
Schulz, R., Sierks, H., Küppers, M., and Accomazzo, A., \enquote{Rosetta
  Fly-by at Asteroid (21) Lutetia: An Overview,} {\em Planetary and Space
  Science\/}, Vol.~66, No.~1, 2012, pp.~2 -- 8.
\newblock \doi{10.1016/j.pss.2011.11.013}.

\end{thebibliography}

\end{document}